\documentclass{Interspeech}
\usepackage{booktabs}
\usepackage{multirow}
\usepackage{graphicx}

\interspeechcameraready 

\title{Discrete Audio Representations for Automated Audio Captioning}

\author[affiliation={1}]{Jingguang}{Tian}
\author[affiliation={1,2}]{Haoqin}{Sun}
\author[affiliation={1}]{Xinhui}{Hu}
\author[affiliation={1}]{Xinkang}{Xu}

\affiliation{}{Hithink RoyalFlush AI Research Institute}{Hangzhou, China}
\affiliation{TMCC, College of Computer Science}{Nankai University}{Tianjin, China}
\email{\{tianjingguang, sunhaoqin, huxinhui, xuxinkang\}@myhexin.com}

\usepackage{comment}

\begin{document}
\maketitle

\begin{abstract}
    % 1000 characters. ASCII characters only. No citations
    Discrete audio representations, termed audio tokens, are broadly categorized into semantic and acoustic tokens, typically generated through unsupervised tokenization of continuous audio representations. However, their applicability to automated audio captioning (AAC) remains underexplored. This paper systematically investigates the viability of audio token-driven models for AAC through comparative analyses of various tokenization methods. Our findings reveal that audio tokenization leads to performance degradation in AAC models compared to those that directly utilize continuous audio representations. To address this issue, we introduce a supervised audio tokenizer trained with an audio tagging objective. Unlike unsupervised tokenizers, which lack explicit semantic understanding, the proposed tokenizer effectively captures audio event information. Experiments conducted on the Clotho dataset demonstrate that the proposed audio tokens outperform conventional audio tokens in the AAC task.
    %Discrete audio representations, known as audio tokens, have received considerable attention due to their capacity to incorporate audio into language models. However, the utilization of audio tokens in the domain of automated audio captioning (AAC) remains relatively unexplored. In this paper, we conduct a comprehensive exploration of the application of audio tokens in AAC models. Our findings reveal that audio tokenization leads to information loss, thereby impairing AAC performance. To mitigate this issue, we propose a novel tokenizer designed to extract audio tokens, enhancing the performance of AAC. Previous tokenizers are typically trained through unsupervised learning and lack explicit semantic knowledge. In contrast, the proposed tokenizer is trained with the objective of audio tagging, enabling it to capture audio event information. Experiments on the Clotho dataset demonstrate that the proposed audio tokens surpass the performance of conventional audio tokens on the AAC task.
\end{abstract}

\section{Introduction}
Automated audio captioning (AAC) is a cross-modal translation task aimed at generating natural language descriptions for audio clips \cite{mei2022automated}. The primary approach employs an encoder-decoder-based deep learning framework, where the audio encoder extracts acoustic features, and the text decoder generates captions based on these features \cite{chen2020audio, gontier2021automated, liu2022leveraging, labb2024conette, wu2024improving}. Recent studies frequently utilize pre-trained audio encoders, such as PANN \cite{PANN}, CNext \cite{pellegrini2023adapting}, or BEATs \cite{chen2023beats}, alongside language models like BERT \cite{devlin2018bert}, BART \cite{lewis2019bart}, and GPT-2 \cite{radford2019language}. These methodologies encode rich information from audio signals into continuous representations. Although continuous representations effectively capture complex audio details, there is growing interest in discrete representations \cite{vashishth2024stab}.

Audio tokenization transforms audio into discrete tokens, enabling seamless integration with large language models (LLMs). Audio tokens can be categorized into two types: acoustic and semantic tokens. Acoustic tokens are produced through neural compression (codecs) to reconstruct the original audio signal, ensuring it remains perceptually identical to listeners. Models like Soundstream \cite{zeghidour2021soundstream}, EnCodec \cite{defossez2022high}, and DAC \cite{kumar2024high} are prominent neural compression models for acoustic tokenization, utilizing an encoder-decoder architecture and a residual vector quantizer (RVQ), where each layer quantizes residuals from the previous layer. In contrast, semantic tokens are derived through representation learning, which aims to maximize the preservation of the audio's semantic information. A common semantic tokenization approach involves selecting a layer from a pre-trained self-supervised learning (SSL) or supervised learning (SL) model and clustering its representations, often using the k-means algorithm, as seen in HuBERT \cite{hsu2021hubert} and WavLM \cite{chen2022wavlm}. RepCodec \cite{huang-etal-2024-repcodec} is a tokenization method that employs parametric networks to extract semantic tokens. Unlike conventional audio codecs that reconstruct the original audio, RepCodec focuses on reconstructing audio representations. However, these audio tokens are typically obtained through unsupervised learning, which inspires an idea: whether they can be further enhanced through supervised training.

Numerous studies have explored the potential of discrete audio tokens as an alternative to continuous representations, particularly in speech-related tasks. The paradigm of LLM-based speech generation, which conditions on text to model audio tokens, has gained prominence due to its ability to produce highly natural-sounding voices \cite{wang2023neural, lajszczak2024base}. Discrete representations have also demonstrated competitive performance in various discriminative tasks \cite{vashishth2024stab, chang23b_interspeech, mousavi24_interspeech, puvvada2024discrete}, including speech recognition, speech translation, speaker recognition, speaker diarization, and speech emotion recognition. However, the application of discrete audio representations in AAC is still in its early stages. EnCLAP \cite{kim2024enclap} has shown that combining EnCodec with continuous audio representations provides a more effective input for pre-trained language models by leveraging the strengths of both discrete and continuous representations. Despite these advancements, the use of audio tokens in AAC systems remains largely unexplored.

In this paper, we explore the integration of discrete audio representations into two representative models, BART Base and GPT-2 XL, for the development of AAC systems. We employ pre-trained BEATs to extract continuous audio representations, which are then discretized using k-means and RepCodec to generate semantic tokens. For acoustic tokens, we utilize EnCodec and DAC. Our comparative analysis reveals that semantic tokens significantly outperform acoustic tokens in the AAC task. However, despite these advancements, we observe that audio discretization degrades the performance of AAC models compared to models that directly utilize continuous representations. To mitigate this issue, we propose a novel approach for deriving semantic tokens through supervised learning. This involves integrating a vector quantization module into pre-trained BEATs and training the model with a multi-label audio tagging loss. Experimental evaluations conducted on the Clotho dataset demonstrate that AAC models utilizing the proposed semantic tokens achieve performance levels that are competitive with those employing continuous representations.
\begin{figure*}[t]
  \centering
  \includegraphics[width=\linewidth,height=6.5cm]{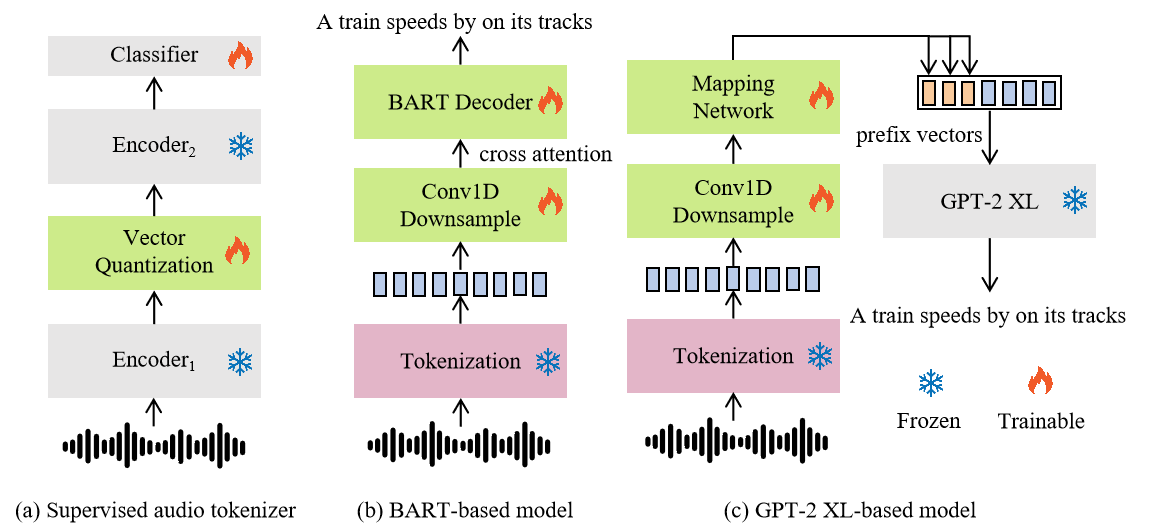}
  \caption{Schematic diagram of the proposed automated audio captioning models utilizing audio tokens.} 
  \label{fig:AAC}
  \vspace{-0.4cm}
\end{figure*}

The remainder of the paper is organized as follows: Section 2 outlines the methodology for constructing AAC systems utilizing discrete audio representations. Section 3 presents the experimental results. Finally, Section 4 concludes the paper.

\section{Method}

This section describes the AAC models that utilize audio tokens, as illustrated in Figure~\ref{fig:AAC}. We construct BART-based and GPT-2 XL-based AAC systems, both utilizing semantic and acoustic tokens. %Detailed descriptions of these tokens and models are presented below.

\subsection{Acoustic tokens}

\subsubsection{EnCodec}

EnCodec is an audio codec based on a convolutional encoder-decoder architecture trained in an end-to-end manner. We employ the pre-trained EnCodec\footnote{https://github.com/facebookresearch/encodec} to process audio at a bitrate of 24 kbps, utilizing 32 codebooks, each containing 1024 code vectors with 128 dimensions. The model transforms 24 kHz audio into discrete token sequences at a frame rate of 75 Hz.
%EnCodec is an audio codec based on a convolutional encoder-decoder architecture with a quantized latent space, trained end-to-end. We utilize the pre-trained EnCodec \footnote{https://github.com/facebookresearch/encodec} to process audio at a bitrate of 24 kbps, employing 32 codebooks, each comprising 1024 code vectors of size 128. The model converts 24 kHz audio waveforms into discrete token sequences at a frequency of 75 Hz. %In our experiments, we avoid using quantized audio codes and instead use the corresponding code vectors as discrete audio representations. Specifically, 75 vectors of dimension 128 are extracted per second of audio.

\subsubsection{DAC}

DAC, an enhanced version of EnCodec, combines advancements in high-fidelity audio generation with improved codebook learning techniques, alongside refined adversarial and reconstruction losses. We use a pre-trained DAC\footnote{https://github.com/descriptinc/descript-audio-codec} with 32 codebooks, each containing 1024 code vectors of size 8, to compress 24 kHz audio into discrete tokens at a frame rate of 75 Hz.
%DAC, an enhanced version of EnCodec, combines advancements in high-fidelity audio generation with improved codebook learning techniques, along with refined adversarial and reconstruction losses. We use a pre-trained DAC \footnote{https://github.com/descriptinc/descript-audio-codec} with 32 codebooks, each containing 1024 code vectors of size 8, to compress 24 kHz audio into discrete tokens at a frame rate of 75 Hz. %Despite the codebook vector dimension being 8, DAC decouples code lookup and code embedding, enabling each second of audio to be encoded into 75 discrete representations of 1024 dimensions.
    
\subsection{Semantic tokens}

\subsubsection{BEATs}

BEATs, a state-of-the-art audio tagging model, is trained using an iterative audio pretraining framework. It transforms 16 kHz audio waveforms into 128-dimensional Mel filter bank features (Fbank) with a 10 ms frame shift. These features are segmented into \(16\times16\) patches and converted into patch embeddings via a linear projection layer, producing 50 patch embeddings per second of audio. The embeddings are then processed by 12 Transformer encoder layers with 768-dimensional hidden states to generate encoded audio representations. We use the output from the ninth layer of the pre-trained BEATs\footnote{https://github.com/microsoft/unilm/tree/master/beats} as continuous audio representations and apply various tokenization methods to obtain semantic tokens.

\subsubsection{K-means}

K-means clustering is a widely used method for semantic token extraction. We first extract continuous audio representations from the training data using BEATs, then determine the number of clusters, \(K\), fit the k-means model (codebook), and obtain \(K\) cluster centroids (code vectors). During the tokenization process, the Euclidean distance between the continuous audio representations and the cluster centroids is calculated. The centroid index with the smallest distance is selected as the discretized token.

\subsubsection{RepCodec}

RepCodec employs an end-to-end neural codec to more effectively preserve information within continuous audio representations. It consists of an encoder, a vector quantizer (VQ) \cite{van2017neural}, and a decoder. The VQ compresses the encoder's output representations into discrete tokens by projecting them onto their closest codebook entries, which are then passed to the decoder to reconstruct the input representations. In this work, we utilize BEATs to extract continuous audio representations for reproducing RepCodec. More details can be found in \cite{huang-etal-2024-repcodec}.

\subsubsection{Supervised audio tokenizer}

Our proposal introduces a supervised audio tokenizer designed to explicitly capture semantic information. This approach contrasts with prevalent audio tokenizers, which are predominantly developed through unsupervised learning methods and thus lack the capacity to acquire explicit knowledge of audio content. Transferring audio event knowledge from audio tagging models—which recognize specific sound events within an audio recording—to AAC results in substantial performance improvements \cite{xu2021investigating}. We hypothesize that a tokenizer trained with audio tagging objectives enhances semantic representation, thereby benefiting AAC. Advanced audio tagging models often undergo a multifaceted process that includes self-supervised pre-training followed by supervised fine-tuning. Rather than training from scratch, we employ a more computationally efficient approach.

Specifically, we split the pre-trained BEATs model into two parts: the first 9 layers form \(Encoder_{1}\), and the last 3 layers form \(Encoder_{2}\). A vector quantization module is then incorporated between the two. The network architecture of this vector quantization module mirrors that of RepCodec \cite{huang-etal-2024-repcodec}. During the training stage, as illustrated in Figure~\ref{fig:AAC} (a), the parameters of \(Encoder_{1}\) and \(Encoder_{2}\) are fixed. Codebook embeddings are updated using an exponentially moving average (EMA), while the encoder and decoder of the vector quantization module are optimized via binary cross-entropy loss. We denote the audio as \(a_{n}\), where \({n}\) is the index of the audio files, and \(f\left ( a_{n} \right )\in \left [ 0,1 \right ]^{S}\) represents the model's output probabilities for \({S}\) sound classes. The label of \(a_{n}\) is denoted as \(l_{n}\in \left\{ 0,1\right\}^{S}\). The binary cross-entropy loss function \(\mathit{L}_{\text{B}}\) is defined as:
\setlength{\abovedisplayskip}{0.1cm}
\setlength{\belowdisplayskip}{0.1cm} 
\begin{equation}
  %\mathit{L}_{BCE}=-\sum_{n=1}^{N} l_{n} \cdot \ln f\left( a_{n} \right) + \left( 1 - l_{n} \right) \cdot \ln \left( 1 - f\left( a_{n} \right) \right)
  \mathit{L}_{\text{B}}=-\sum_{n=1}^{N} \left( l_{n} \cdot \ln f\left( a_{n} \right) + \left( 1 - l_{n} \right) \cdot \ln \left( 1 - f\left( a_{n} \right) \right) \right)
\end{equation}
where \({N}\) is the number of training samples. It is worth noting that our supervised tokenizer training scheme is adaptable to any pre-trained audio tagging model. Empirical evidence shows that selecting higher layers for splitting audio tagging models yields better results. After training, \(Encoder_{1}\), the encoder of the vector quantization module, and the vector quantizer form a pipeline for converting audio waveforms into semantic tokens.

\subsection{AAC models}
\subsubsection{BART-based model}

The BART-based AAC model employs an encoder-decoder architecture, as shown in Figure~\ref{fig:AAC} (b). The encoder utilizes a single-layer 1D convolution with a kernel size of 3 and a stride of 3, primarily aimed at achieving a threefold reduction in the input sequence length. A 6-layer BART Base Transformer decoder is utilized for caption generation, with the pretrained BART\footnote{https://huggingface.co/facebook/bart-base} text tokenizer applied during training and the decoder weights initialized randomly. The BART decoder cross-attends to the encoder’s output while self-attending to previous caption tokens in an autoregressive manner. Given the input audio representations \(\boldsymbol{x}= x_{1},x_{2},\cdot \cdot \cdot x_{n}\), and their corresponding captions \(\boldsymbol{y}= y_{1},y_{2},\cdot \cdot \cdot y_{m}\), the model's training objective is to minimize cross-entropy loss:
\setlength{\abovedisplayskip}{0.1cm}
\setlength{\belowdisplayskip}{0.1cm}
\begin{align}
 \mathit{L}_{\text{BART}}= -\frac{1}{m}\sum_{t=1}^{m}log\mathit{p}\left (y_{t}|y_{1:t-1},\boldsymbol{x} \right )
\end{align}
where \(y_{t}\) is the \(t^{th}\) token in the caption.

\subsubsection{GPT-2 XL-based model}

The GPT-2 XL-based AAC model employs a decoder-only framework, as shown in Figure~\ref{fig:AAC} (c), inspired by the work of \cite{kim2023prefix}. The input representation sequence undergoes a \(3\times \) downsampling via a Conv1D layer, analogous to the BART-based model. To handle varying audio representation lengths, we concatenate a fixed number of learnable embeddings with the downsampled features and feed them into a mapping network consisting of a single Transformer layer with 8 heads and an embedding dimension of 1,600. From the output embeddings of the mapping network, we discard the portion corresponding to the downsampled features and retain only the portion corresponding to the learnable embeddings as prefix vectors for GPT-2 XL. Subsequently, the GPT-2 XL model generates captions conditioned on prefix vectors \(\boldsymbol{p}= p_{1},p_{2},\cdot \cdot \cdot p_{k}\), where k is the prefix length. In our experiments, k is set to 50. During training, the pre-trained GPT-2 XL\footnote{https://huggingface.co/openai-community/gpt2-xl} model remains frozen, while the Conv1D and mapping network are updated to learn prefix mappings comprehensible to the GPT-2 XL. The model is trained by cross-entropy loss, calculated solely for the tokens corresponding to the captions:
\setlength{\abovedisplayskip}{0.1cm}
\setlength{\belowdisplayskip}{0.1cm}
\begin{align}
  \mathit{L}_{\text{GPT-2 XL}}= -\frac{1}{m}\sum_{t=1}^{m}log\mathit{p}\left (y_{t}|\boldsymbol{p},y_{1:t-1} \right)
\end{align}
\section{Experiment}

This section provides a detailed description of the datasets, experimental setup, and evaluation metrics used in our study.

\subsection{Datasets}

AudioSet \cite{gemmeke2017audio} is an audio dataset comprising 527 sound classes and approximately 2 million 10-second audio clips. The dataset is divided into three subsets: an unbalanced subset containing 2,042,985 clips, a balanced subset with 22,176 clips, and an evaluation set comprising 20,383 clips. We utilize the AudioSet dataset to train audio tokenizers for extracting semantic tokens.

Clotho \cite{drossos2020clotho} consists of 15 to 30-second audio clips, each paired with five captions containing 8 to 20 words. For our experiments, we use the Clotho v2 dataset, which comprises 3,839 development samples, 1,045 validation samples, and 1,045 evaluation samples. The development set is used to train the AAC models, the validation set for parameter tuning, and the evaluation set to assess the model's performance.

\subsection{Experimental setup}

\subsubsection{Tokenizer model}

We classify Clotho as in-domain data and AudioSet as out-of-domain data to investigate the domain effect on the tokenizer. The k-means model is trained on the Clotho development set and the balanced subset of AudioSet, with \(K\) set to 1,024. The RepCodec model is trained on the Clotho development set and the unbalanced subset of AudioSet. Due to the absence of sound labels in Clotho, the supervised audio tokenizer is exclusively trained on the unbalanced subset of AudioSet. For RepCodec and the supervised audio tokenizer, we employ two types of quantizers—conventional VQ \cite{van2017neural} and RVQ \cite{zeghidour2021soundstream}—both utilizing codebooks of size 1,024. Notably, RVQ uses two codebooks; with one, it is equivalent to VQ. Both models are trained for 100K steps with a batch size of 64, using the AdamW optimizer and a constant learning rate of 1e-5.

During the discretization process, we forgo using quantized audio codes, opting instead for the corresponding code vectors as discrete audio representations, due to their slight performance advantage. For tokenizers with multiple codebooks, the corresponding code vector sequences are summed to form the final quantized representation.
%We classify Clotho as in-domain data and AudioSet as out-of-domain data to investigate the domain effect on the tokenizer. The k-means model is trained on the Clotho development set and the balanced subset of AudioSet, with \(K\) set to 1,024. The RepCodec model is trained on the Clotho development set and the unbalanced subset of AudioSet. Due to the absence of sound labels in Clotho, the supervised audio tokenizer is exclusively trained on the unbalanced subset of AudioSet. For RepCodec and the supervised audio tokenizer, we employ two types of quantizers—conventional VQ \cite{van2017neural} and RVQ \cite{zeghidour2021soundstream}—both utilizing a codebook size of 1,024. RVQ uses two codebooks, and with one, it is equivalent to VQ. Both models are trained for 100K steps with a batch size of 64, using the AdamW optimizer and a constant learning rate of 1e-5.
%During the discretization process, we forgo using quantized audio codes, opting instead for the corresponding code vectors as discrete audio representations due to their slight performance advantage. For tokenizers with multiple codebooks, the corresponding code vector sequences are summed to form the final quantized representation.
%
%
\begin{table*}[th]
\caption{The performance of AAC models on the Clotho evaluation set. RepCodec-VQ/RVQ and Supervised-VQ/RVQ denote the RepCodec and supervised audio tokenizer with VQ/RVQ, respectively. ID indicates training the tokenizer with in-domain data, while OD indicates training with out-of-domain data. \(^{*}\) indicates that the results are sourced from \cite{kim2024enclap}.}
\label{tab:results}
\centering
\fontsize{8.5}{12}\selectfont
\begin{tabular}{cccccccccccccc}
\hline
\multirow{3}{*}{\textbf{Tokenizer}} & \multicolumn{6}{c}{\textbf{BART-based AAC model}}                                                         & \multicolumn{6}{c}{\textbf{GPT-2 XL-based AAC model}}                                                     & \textbf{ENCLAP}\(^{*}\)                   \\
\cmidrule(r){2-7} \cmidrule(r){8-13} \cmidrule(r){14-14}
                           & \multicolumn{2}{c}{SPIDEr (↑)} & \multicolumn{2}{c}{FENSE (↑)} & \multicolumn{2}{c}{\#Words (↑)} & \multicolumn{2}{c}{SPIDEr (↑)} & \multicolumn{2}{c}{FENSE (↑)} & \multicolumn{2}{c}{\#Words (↑)} & SPIDEr (↑)               \\ 
                           \cmidrule(r){2-3} \cmidrule(r){4-5} \cmidrule(r){6-7} \cmidrule(r){8-9} \cmidrule(r){10-11} \cmidrule(r){12-13} \cmidrule(r){14-14}
                           & ID             & OD            & ID            & OD            & ID             & OD             & ID                 & OD        & ID            & OD            & ID             & OD             & \multirow{10}{*}{0.294} \\ \cline{1-13}
K-means                    & 0.285         & 0.267        & 0.483        & 0.474        & 436            & 457            & 0.256             & 0.256    & 0.474        & 0.463        & 725            & 757            &                          \\
RepCodec-VQ                & 0.288         & 0.275        & 0.483        & 0.471        & 459            & 440            & 0.265             & 0.256    & 0.471        & 0.469        & 561            & 715            &                          \\
RepCodec-RVQ               & 0.294         & 0.292        & 0.496        & 0.487        & 475            & 488            & \textbf{0.279}    & 0.268    & 0.485        & 0.477        & 692            & 658            &                          \\
Supervised-VQ              & —              & 0.285        & —             & 0.479         & —              & 425            & —                  & 0.268    & —             & 0.477        & —              & 684            &                          \\
Supervised-RVQ             & —              & 0.294        & —             & 0.490        & —              & 474            & —                  & 0.277    & —             & 0.484        & —              & 693            &                          \\
EnCodec                    & —              & 0.085        & —            & 0.217        & —              & 98             & —                  & 0.129    & —             & 0.284        & —              & 448            &                          \\
DAC                        & —              & 0.143        & —            & 0.322        & —              & 235            & —                  & 0.154    & —             & 0.324        & —              & 487            &                          \\  \cline{1-13}
FBANK                      & \multicolumn{2}{c}{0.125}     & \multicolumn{2}{c}{0.280}    & \multicolumn{2}{c}{152}         & \multicolumn{2}{c}{0.076}     & \multicolumn{2}{c}{0.214}    & \multicolumn{2}{c}{45}          &                          \\
BEATs                      & \multicolumn{2}{c}{\textbf{0.299}}     & \multicolumn{2}{c}{\textbf{0.503}}    & \multicolumn{2}{c}{\textbf{493}}         & \multicolumn{2}{c}{0.270}     & \multicolumn{2}{c}{\textbf{0.486}}    & \multicolumn{2}{c}{\textbf{837}}         &                          \\ \hline
\end{tabular}
\vspace{-0.4cm}
\end{table*}
\subsubsection{AAC model}

We explore five discrete audio representations and employ two continuous audio representations (FBANK and BEATs) for comparative analysis in building AAC models. All input audio representations are extracted from 30-second segments, with zero-padding applied to shorter audio clips. Since different input audio representations have varying dimensions, the input dimension of Conv1D is adjusted accordingly. Captions are converted to lowercase, and punctuation is removed. The model is trained for 20 epochs on the Clotho development set with a batch size of 16, using the AdamW optimizer, an initial learning rate of 2e-5 with linear decay, and a weight decay of 3e-4. During inference, beam search with a beam size of 3 is utilized.
%We apply SpecAugment \cite{park19e_interspeech} to the input audio representations, using time and frequency drop widths of 64.
\subsection{Metrics}

To evaluate the performance of the models, we employ SPIDEr \cite{liu2017improved}, FENSE \cite{zhou2022can}, and the unique word vocabulary of captions (referred to as \#Words). While SPIDEr, though valuable for comparative analysis with existing literature, exhibits high sensitivity to n-gram matching, FENSE is a metric based on two pre-trained models that compares sentence embeddings rather than n-grams. The unique word vocabulary provides insights into the diversity of the model's outputs.

\subsection{Results}

We present the results on the Clotho evaluation set in Table~\ref{tab:results}. EnCodec and DAC perform significantly worse than k-means, RepCodec, and the proposed supervised audio tokenizer, as they are designed to reconstruct original audio, focusing on acoustic details while neglecting crucial semantic information. BEATs features, rich in high-level semantics, are effectively discretized by k-means, RepCodec, and the proposed supervised audio tokenizer, preserving their semantic information and making them more suitable for AAC. The bottom row of Table~\ref{tab:results} shows that using continuous audio representations from BEATs in AAC models outperforms using discrete representations from k-means, RepCodec, and the proposed supervised audio tokenizer. This indicates that discretizing audio representations leads to information loss. However, the proposed supervised audio tokenizer achieves the best performance among tokenization methods on out-of-domain training data, nearly matching the performance of continuous audio representations. This suggests that supervised learning with audio tagging enables the tokenizer to capture knowledge of audio events, thereby mitigating information loss during discretization.

Further analysis reveals that the GPT-2 XL-based AAC model underperforms compared to the BART-based model in both SPIDEr and FENSE metrics, although it generates more diverse captions. Given Clotho's limited size, we hypothesize that the GPT-2 XL-based AAC model is more prone to overfitting in low-resource settings, despite retaining the robust language expressivity of the LLM. In most scenarios, the AAC model using discrete audio representations outperforms the FBANK-based model. Moreover, our BART-based model, which uses only discrete audio representations, performs comparably to the ENCLAP model that leverages both discrete and continuous audio representations. These findings highlight the significant potential of discrete audio representations for AAC.
\begin{figure}[t]
  \centering
  \vspace{-0.25cm}
  \includegraphics[width=7.0cm,height=5.0cm]{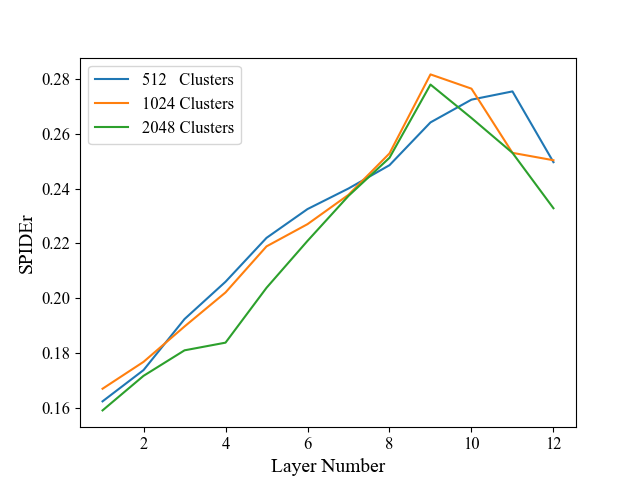}
  \caption{Performance of k-means applied to the BART-based AAC model across different layers and cluster numbers.}
  \label{fig:kmeans}
  \vspace{-0.65cm}
\end{figure}
\subsection{Ablation studies}

Employing k-means clustering as a case study, we investigate how layer selection within the BEATs model and codebook size (number of clusters) affect semantic token performance, as depicted in Figure~\ref{fig:kmeans}. The results indicate that features from the upper layers are more suitable for AAC, and the optimal number of layers and clusters is interdependent. Additionally, we explore several pivotal factors, including the influence of training data domains and the number of codebooks used by the quantizer for generating semantic tokens, as illustrated in Table~\ref{tab:results}. For k-means and RepCodec, in-domain training data outperforms out-of-domain data. Moreover, for RepCodec and the proposed supervised audio tokenizer, increasing the number of codebooks in the quantizer enhances performance.
\vspace{-0.1cm}

\section{Conclusion}

This paper presents the first systematic study of discrete audio representations for AAC. Our findings indicate that semantic tokens significantly outperform acoustic tokens in this context. Moreover, the discretization of continuous audio representations leads to information loss, diminishing AAC performance. To mitigate this, we develop a supervised audio tokenizer to extract semantic tokens, leveraging an audio tagging objective for training. This method incorporates explicit semantic knowledge to compensate for the information loss due to discretization. Experimental results show that the proposed supervised audio tokenizer surpasses existing methods and achieves comparable results to continuous representations in the AAC task.
%However, a performance gap persists between discrete and continuous audio representations, highlighting the need for further research.
%This paper investigates how discrete audio representations can be leveraged for the AAC task. Our study demonstrates that semantic tokens significantly outperform acoustic tokens in AAC tasks. Additionally, the discretization of continuous audio representations results in information loss, which reduces AAC performance. 
%
%To alleviate this issue, we utilize an audio tagging objective to train the audio tokenizer, thereby incorporating additional semantic information to compensate for the information loss caused by discretization.
%
%The comprehensive experimental results indicate that the proposed supervised audio tokenizer surpasses all existing audio tokenization methods in AAC tasks.
%However, a performance gap persists between discrete and continuous audio representations, highlighting the need for further research.
%
\bibliographystyle{IEEEtran}
\bibliography{mybib}

\end{document}